\documentclass{elsarticle}

\usepackage{amsmath}
\usepackage{bm}
\usepackage{tikz}
\usepackage{pgfplots}

\usepackage{hyperref}

\newproof{pf}{Proof}

\journal{arXiv.org} 

\newtheorem{pr}{Proposition}
\newtheorem{as}{Assumption}
\newtheorem{rem}{Remark}

\newcommand{\grad}{\mathop{\rm grad}\nolimits}
\renewcommand{\div}{\mathop{\rm div}\nolimits}
\newcommand{\const}{\mathop{\rm const}\nolimits}

\graphicspath{{figs/}}

\begin{document}

\begin{frontmatter}

\title{Decoupling schemes for predicting compressible fluid flows}

\author[nsi,rudn]{Petr N. Vabishchevich\corref{cor}}
\ead{vabishchevich@gmail.com}

\address[nsi]{Nuclear Safety Institute, Russian Academy of Sciences, 52, B. Tulskaya, Moscow, Russia}
\address[rudn]{Peoples' Friendship University of Russia (RUDN University), 6 Miklukho-Maklaya St., 117198 Moscow, Russia}

\cortext[cor]{Corresponding author}

\begin{abstract}
Numerical simulation of compressible fluid flows is performed using the Euler equations.
They include the scalar advection equation for the density, the vector advection equation for the velocity
and a given pressure dependence on the density. An approximate solution of an initial--boundary value problem
is calculated using the finite element approximation in space.
The fully implicit two--level scheme is used for discretization in time. 
Numerical implementation is based on Newton's method.
The main attention is paid to fulfilling conservation laws for the mass and total mechanical energy
for the discrete formulation. Two--level schemes of splitting by physical processes are employed for numerical solving
problems of barotropic fluid flows.
For a transition from one time level to the next one, an iterative process is used, where at each iteration
the linearized scheme is implemented via solving individual problems for the density and velocity.
Possibilities of the proposed schemes are illustrated by numerical results for a two--dimensional model problem 
with density perturbations.
\end{abstract}


\end{frontmatter}

\section{Introduction}

Applied models of continuum mechanics~\cite{Batchelor,LandauLifshic1986}
are based on conservation laws for the mass, momentum and energy.
The transport of scalar and vector quantities due to advection determines a mathematical form of
conservation laws \cite{Godunov,LeVeque}.
In addition, some parameters of a flow have the positivity property (monotonicity).
Such important properties of the differential problem of continuum mechanics must be inherited
in a discrete problem~\cite{Anderson,Wesseling}.

Flows of ideal fluids are governed by the Euler equations, whereas the Navier--Stokes equations are applied 
to describing viscous flows. Mathematical problems of validation of such models are considered, for example,
in the books~\cite{lions1,lions2}. When discussing the existence of solutions in various Sobolev spaces,
the principal problems of the positivity (non--negativity) of the fluid density are also should be highlighted.
Such a consideration is also carried out (see, for instance, \cite{feireisl2016mathematical})
at the discrete level for various approximations in time and space.

In computational fluid dynamics, the most important problems are associated with two contradictory requirements.
Namely, it is necessary to construct monotone approximations for advective terms and to fulfil conservation laws. 
The construction of monotone approximations is discussed in many papers 
(see, e.g., \cite{kulikovskii2000mathematical,HundsdorferVerwer2003,Kuzmin}).  
In~\cite{MortonKellogg1996,SamarskiiVabischevich1999a}, standard linear approximations are considered
for the basic problems of continuum mechanics (convection--diffusion problems).

For discretization in space, conservative approximations are constructed on the basis of using the conservative 
(divergent) formulation of continuum mechanics equations.
This approach is most naturally implemented using integro--interpolation method (balance method) for
regular and irregular grids~\cite{Samarskii1989}, and in the control method volume~\cite{LeVeque,Versteeg}.
Nowadays, the main numerical technique to solve applied problems is the finite element method~\cite{Guermond,Larson}.
It is widely used in computational fluid dynamics~\cite{Donea,Zienkiewicz}, too.

Discretizations in time for computational fluid dynamics are often constructed using
explicit schemes that have strong restrictions on a time step in sense of stability.
Moreover, explicit schemes have similar restrictions on the monotonicity of an approximate solution. 
So, it is more natural to focus on implicit schemes.
To solve boundary value problems for partial differential equations, two--level schemes are widely used
\cite{HundsdorferVerwer2003,Ascher2008,LeVeque2007} ($\theta$--method, schemes with weights).
For linear problems, a study of discretizations in time can be based on the general theory of stability (well--posedness)
for operator--difference schemes~\cite{Samarskii1989,SamarskiiMatusVabischevich2002}. 
In particular, it is possible to apply unimprovable (coinciding necessary and sufficient) stability conditions,
which are formulated as operator inequalities in finite--dimensional Hilbert spaces.

In the present work, an initial--boundary value problem is considered for the Euler equations describing 
barotropic fluid flows (Section 2), which are conservation laws for the mass, momentum, and total mechanical energy.
Discretization in space is performed (Section 3) using standard Lagrange elements for the density and cartesian 
velocity components. To evaluate an approximate solution at a new time level, the fully implicit scheme is employed.
For the approximate solution, the mass conservation law holds and an estimate for the dissipation 
of the total mechanical energy is fulfilled. The fully implicit scheme is not convenient for numerical implementation.
The solution at the new time level is determined from a system of coupled nonlinear equations for the density and velocity.
A decoupling scheme is proposed in Section 4, which refers to the class of linearized schemes of splitting by physical
processes~\cite{Marchuk1990,Vabishchevich2014}.
Linearization is carried out over the field of advective transport in such a way that at each time level we
solve individual problems for the density and velocity.
Possibilities of the proposed schemes are illustrated by the results of numerical solving a model
two--dimensional problem with a perturbation of the fluid density being initially at rest (Section 5).
To solve numerically the nonlinear discrete problem at the new time level, the Newton method is used.
In the above calculations, a small number of iterations (two or three) is sufficient for the process convergence.
The influence of the grid size in space and time is investigated.
It was observed that decreasing of the time step results in the monotonization of the numerical solution.
The main result of the paper is the proof of the robustness of the linearized decoupling scheme.
The scheme involves separate solving standard advection problems for the density and velocity
and demonstrates high iteration convergence.  Such an approach can be used for other problems of continuum mechanics, 
e.g., for numerical solving initial--boundary value problems for the Navier--Stokes equations.

\section{Mathematical models} 

An initial--boundary value problem is considered for describing barotropic fluid flows.
The system of equations includes the scalar advection equation for the density 
and the vector advection equation for the velocity with a given pressure dependence on the density.
The conservation laws for the mass, momentum, and total mechanical energy are discussed.

\subsection{Barotropic fluid} 

The continuity equation in a bounded domain $\Omega$ has the form
\begin{equation}\label{1}
 \frac{\partial \varrho}{\partial t} + \div(\varrho \bm u) = 0 ,
 \quad \bm x \in \Omega ,
 \quad 0 < t \leq T ,
\end{equation} 
where $\varrho(\bm x, t) > 0$ is the density and $\bm u (\bm x, t)$ is the velocity.
The momentum equation is written in the conservative form
\begin{equation}\label{2}
 \frac{\partial }{\partial t} (\varrho \bm u) + \div(\varrho \bm u \otimes \bm u) + \grad p = 0 , 
 \quad \bm x \in \Omega ,
 \quad 0 < t \leq T ,
\end{equation} 
where $p(\bm x, t)$ is the pressure. The considered fluid is assumed to be barotropic, i.e.,
we have a known dependence of the pressure on the density
$p = p(\varrho)$, ${\displaystyle \frac{d p}{d \varrho} > 0}$.

Assume that the domain boundaries are rigid and so, the impermeability condition is imposed:
\begin{equation}\label{3}
 (\bm u \cdot \bm n) = 0, 
 \quad \bm x \in \partial \Omega . 
\end{equation}
Initial conditions for the density and velocity are also specified:
\begin{equation}\label{4}
 \varrho(\bm x, 0) = \varrho^0(\bm x) ,
 \quad \bm u (\bm x, 0) = \bm u^0(\bm x) ,
 \quad \bm x \in \Omega .
\end{equation}
The initial--boundary value problem (\ref{1})--(\ref{4}) describes transient flows of
an ideal barotropic fluid.

The direct integration of the continuity equation (\ref{1}) over the domain $\Omega$ taking into account 
the boundary condition (\ref{3}) results in the mass conservation law
\begin{equation}\label{5}
 m (t) = m(0) ,
 \quad m(t) = \int_{\Omega}  \varrho(\bm x, t) d \bm x .
\end{equation}
In the Hilbert space $L_2(\Omega)$, we define the scalar product and norm in the standard way:
\[
  (w,u) = \int_{\Omega} w({\bm x}) u({\bm x}) d{\bm x},
  \quad \|w\| = (w,w)^{1/2} .
\]
In a similar way, the space of vector functions $\bm L_2(\Omega)$ is defined.
If the density is non--negative, the conservation law for the mass can be written as
\[
 \|\varrho^{1/2}\| = \|\varrho_0^{1/2}\| .
\]
This relation can be treated as an a priori estimate for $\varrho^{1/2}$ in $L_2(\Omega)$.

The equation (\ref{2}) directly expresses the conservation law for the momentum.
Integrating this equation over $\Omega$, we obtain
\[
 \int_{\Omega} \frac{\partial }{\partial t} (\varrho \bm u) d \bm x + \int_{\partial \Omega} p(\varrho) \bm n d \bm x = 0.
\]   
Thus, we have 
\begin{equation}\label{6}
 \bm I (t) = \bm I(0) - \int_{\partial \Omega} p(\varrho) \bm n d \bm x ,
 \quad \bm I(t) =  \int_{\Omega} \varrho \bm u d \bm x .
\end{equation}

Multiplying by $\bm u$ and taking into account equation (\ref{1}), rewrite equation (\ref{2}) as
\[
 \frac{1}{2} \frac{\partial }{\partial t} (\varrho |\bm u |^2) +
 \frac{1}{2} \div (\varrho |\bm u |^2 \bm u) + \div (p(\varrho) \bm u) - p \div \bm u = 0 . 
\] 
Integration over the domain $\Omega$, in view of (\ref{3}), leads to
\begin{equation}\label{7}
 \frac{1}{2} \frac{d }{d t} \int_{\Omega} \varrho |\bm u |^2 d \bm x - 
 \int_{\Omega} p(\varrho) \div \bm u \, d \bm x = 0 .
\end{equation} 
The second term in (\ref{7}) is expressed from the renormalized equation of continuity.
Define the pressure potential $\varPi(\varrho)$ from the equation
\begin{equation}\label{8}
 \varrho \frac{d \varPi}{d \varrho} - \varPi(\varrho) = p(\varrho) .
\end{equation} 
In particular, for an ideal fluid, we have
\begin{equation}\label{9}
 p(\varrho)  = a\varrho^\gamma, 
 \quad \varPi (\varrho) = a \frac{\varrho^\gamma}{\gamma -1} ,
 \quad a = \const > 0,
 \quad \gamma > 1 . 
\end{equation} 

From the continuity equation (\ref{1}), we have
\begin{equation}\label{10}
 \frac{\partial \varPi }{\partial t} + \div(\varPi  \bm u) + p(\varrho) \div \bm u = 0 ,
 \quad 0 < t \leq T . 
\end{equation}
Integration of the renormalized equation of continuity (\ref{10}) results in the expression
\[
 \frac{d }{d t} \int_{\Omega} \varPi \, d \bm x +
 \int_{\Omega} p(\varrho) \div \bm u \, d \bm x = 0 .
\]
Adding this equality to (\ref{7}), we get
\begin{equation}\label{11}
 \frac{d }{d t}  \int_{\Omega}\left ( \frac{1}{2} \varrho |\bm u |^2 
 + \varPi(\varrho) \right ) d \bm x = 0 .
\end{equation} 
We arrive to the conservation law for the total mechanical energy:
\begin{equation}\label{12}
 E(t) = E(0),
 \quad E(t) =  \int_{\Omega}\left ( \frac{1}{2} \varrho |\bm u |^2 
 + \varPi(\varrho) \right ) d \bm x .
\end{equation} 

The equations (\ref{5}), (\ref{6}) and (\ref{12}) are the basic conservation laws for
of the problem (\ref{1})--(\ref{4}). 

\subsection{Operator--differential formulation}

For the convenience of consideration, we introduce operators of advective (convective) transport for
the system of Euler equations. The advection operator $\mathcal{A} = \mathcal{A}(\bm u)$  
in the divergent form is written as follows:
\begin{equation}\label{13}
 \mathcal{A}(\bm u) \varphi  =  \div  (\bm u \, \varphi ) . 
\end{equation}
Assuming that the boundary condition (\ref{3}) is satisfied for the velocity $\bm u$,
we obtain
\begin{equation}\label{14}
 (\mathcal{A}(\bm u) \varphi, 1) = 0 .
\end{equation}  

The continuity equation (\ref{1}) can be written in the form of an operator--differential equation:
\begin{equation}\label{15}
 \frac{d \varrho }{d t} + \mathcal{A}(\bm u) \varrho  = 0,
 \quad 0 < t \leq T, 
\end{equation}
where the notation $\varrho(t) = \varrho (\bm x,t)$ is used. 
Similarly, equation (\ref{2}) is written in the form
\begin{equation}\label{16}
 \frac{d  }{d t} (\varrho \bm u) + \mathcal{A}(\bm u) (\varrho \bm u) + \grad p(\varrho) = 0,
 \quad 0 < t \leq T.  
\end{equation} 
For the system of equations (\ref{15}), (\ref{16}) with a prescribed dependence $p(\varrho)$,
we consider the Cauchy problem, where the initial conditions (see (\ref{4})) have the form
\begin{equation}\label{17}
 \varrho(0) = \varrho^0,
 \quad \bm u (0) = \bm u^0 .
\end{equation}

For the considered problem (\ref{15})--(\ref{17}), the key point is the property (\ref{14}) 
for the advection operator written in the divergent form.

\section{Implicit two--level scheme}
 
To solve numerically the initial--boundary value problem for the Euler equations, we use
the fully implicit (backward Euler) scheme for time--stepping with finite element discretizations in space.
The problems of fulfilment of the conservation laws at the discrete level are discussed.

\subsection{Discretization in space} 

To solve numerically the problem (\ref{15})--(\ref{17}) (or (\ref{15})--(\ref{19})), we employ finite element 
discretizations in space (see, e.g., \cite{Thomee2006,brenner2008mathematical}). 
For (\ref{13}), we define the bilinear form
\[
 a(\varphi , \psi ) = \int_{\Omega } \div (\bm u \, \varphi ) \, \psi  \, d {\bm x} .
\] 
Define the subspace of finite elements $V^h \subset H^1(\Omega)$ and the discrete operator $A = A(\bm u)$ as
\[
 (A(\bm u) \varphi , \psi ) = a(\varphi , \psi ),
\quad \forall \ w, u \in V^h . 
\]
Similarly to (\ref{14}), we have
\begin{equation}\label{18}
 (A (\bm u) \varphi , 1 ) = 0 .
\end{equation} 

For two-- and three--dimensional vector quantities, the coordinate--wise representation is employed: 
$\bm u = (u_1, ...., u_d)^T, \ d = 2,3$.
For a simple specification of the boundary conditions (\ref{3}), assume
that separate parts of the boundary of the computational domain are parallel to the coordinate axes.
A finite element approximation is used for the individual components of the vector 
$u_i \in V^h, \ i = 1, ..., d$. 

After constructing discretizations in space, we arrive at the Cauchy problem for the system of semi--discrete
operator equations in the corresponding finite--dimensional space, namely, 
we have the Cauchy problem for the system of ordinary differential equations.
For instance, for (\ref{15})--(\ref{17}), we put into the correspondence the problem
\begin{equation}\label{19}
 \frac{d \varrho }{d t} + A(\bm u) \varrho  = 0,
\end{equation} 
\begin{equation}\label{20}
 \frac{d  }{d t} (\varrho \bm u) + A(\bm u) (\varrho \bm u) + \grad p(\varrho) = 0,
 \quad 0 < t \leq T.  
\end{equation} 
\begin{equation}\label{21}
 \varrho(0) = \varrho_0,
 \quad \bm u (0) = \bm u_0 .
\end{equation} 
Here $\varrho_0 = P \varrho^0$, $\bm u_0 = P \bm u^0$ with $P$ denoting $L_2$--projection onto $V^h$.

The solution of the problem (\ref{18})--(\ref{21}) satisfies the same system of conservation laws 
as the solution of the problem (\ref{14})--(\ref{17}) (see (\ref{5}), (\ref{6}), (\ref{12})). 

\subsection{Discretization in time} 

Let $\tau$ be a step of a uniform, for simplicity, grid in time such that 
$\varphi_n = \varphi(t_n), \ t_n = n \tau$, $n = 0,1, ..., N, \ N\tau = T$. 
To construct and study time--stepping schemes, the main attention is given 
to the fulfillment of the corresponding conservation laws (a priori estimates).
Such an important problem as the positivity (non--negativity) of the density at each time level
requires a more in--depth study and so, it is not considered in the present work.

To solve numerically the problem (\ref{19})--(\ref{21}), the fully implicit scheme is applied.
In this case, the approximate solution at the new time level is determined from
\begin{equation}\label{22}
 \frac{\varrho_{n+1} - \varrho_{n} }{\tau} + A(\bm u_{n+1}) \varrho_{n+1}  = 0,
\end{equation} 
\begin{equation}\label{23}
 \frac{\varrho_{n+1} \bm u_{n+1} - \varrho_{n} \bm u_{n}  }{\tau } + 
 A(\bm u_{n+1}) (\varrho_{n+1} \bm u_{n+1}) + \grad p(\varrho_{n+1}) = 0, 
 \quad n = 0, 1, ..., N-1 ,
\end{equation} 
using the prescribed (see (\ref{21})) value $\varrho_0, \bm u_0$.
The basic properties of the approximate solution are related to the fulfillment of the conservation laws
for the mass and total energy. To simplify our investigation, assume that the density is positive.

\begin{as}\label{as-1}
At each time level $\varrho_{n}>0$, $n = 0, 1, ..., N$.
\end{as}

In view of (\ref{18}), integration of equation (\ref{22}) over the domain leads to
\begin{equation}\label{24}
 (\varrho_{n+1}, 1) = (\varrho_{n}, 1),
 \quad n = 0, 1, ..., N-1 .
\end{equation}
The equality (\ref{24}) is a discrete analog the mass conservation law (\ref{5}).
For the momentum conservation law (\ref{6}), we put into the correspondence the equality
\begin{equation}\label{25}
 (\varrho_{n+1} \bm u_{n+1}, 1) = (\varrho_{n} \bm u_{n}, 1)
 - \tau \int_{\partial \Omega} p(\varrho_{n+1})  \bm n d \bm x ,
 \quad n = 0, 1, ..., N-1 , 
\end{equation} 
which has been obtained by integrating equation (\ref{23}). 

An estimate for the total mechanical energy can be established, e.g., following the work~\cite{feireisl2016mathematical}. 
Multiplying equation (\ref{23}) by $\bm u_{n+1}$ and integrating it over $\Omega$, we arrive at
\begin{equation}\label{26}
\begin{split}
 \left ( \frac{\varrho_{n+1} \bm u_{n+1} - \varrho_{n} \bm u_{n}  }{\tau } , \bm u_{n+1} \right )  & + 
 (A(\bm u_{n+1}) (\varrho_{n+1} \bm u_{n+1}), \bm u_{n+1} ) \\
 & + (\grad p(\varrho_{n+1}), \bm u_{n+1} ) = 0 . 
\end{split}
\end{equation}  
For the first term, we have
\[
\begin{split}
 \frac{\varrho_{n+1} \bm u_{n+1} - \varrho_{n} \bm u_{n}  }{\tau } \bm u_{n+1} 
 & = \frac{1}{2} \frac{\varrho_{n+1} |\bm u_{n+1}|^2 - \varrho_{n} |\bm u_{n}|^2  }{\tau } \\
 & + \frac{1}{2} \frac{\varrho_{n+1} |\bm u_{n+1}|^2 + \varrho_{n} |\bm u_{n}|^2 - 2 \varrho_{n} \bm u_{n} \bm u_{n+1} }{\tau } \\
 & =  \frac{1}{2} \frac{\varrho_{n+1} |\bm u_{n+1}|^2 - \varrho_{n} |\bm u_{n}|^2  }{\tau } \\
 & + \frac{1}{2} \frac{\varrho_{n+1} - \varrho_{n} }{\tau }|\bm u_{n+1}|^2
 + \frac{1}{2} \varrho_{n}  \frac{ |\bm u_{n+1} - \bm u_{n}|^2 }{\tau } \\
 & \leq  \frac{1}{2} \frac{\varrho_{n+1} |\bm u_{n+1}|^2 - \varrho_{n} |\bm u_{n}|^2  }{\tau } 
 + \frac{1}{2} \frac{\varrho_{n+1} - \varrho_{n} }{\tau }|\bm u_{n+1}|^2 .
\end{split}
\] 
In view of this, from (\ref{26}), we obtain
\begin{equation}\label{27}
\begin{split}
 \frac{1}{2} \left ( \frac{\varrho_{n+1} |\bm u_{n+1}|^2 - \varrho_{n} |\bm u_{n}|^2  }{\tau } , 1 \right ) 
 & + \frac{1}{2} \left (\frac{\varrho_{n+1} - \varrho_{n} }{\tau }, |\bm u_{n+1}|^2 \right ) \\
 & + (A(\bm u_{n+1}) (\varrho_{n+1} \bm u_{n+1}), \bm u_{n+1} ) \\
 & + (\grad p(\varrho_{n+1}), \bm u_{n+1} ) \leq  0 . 
\end{split}
\end{equation} 

From (\ref{22}) and the definition of the operator $A$, we have 
\[
\begin{split}
\frac{1}{2} \left (\frac{\varrho_{n+1} - \varrho_{n} }{\tau }, |\bm u_{n+1}|^2 \right ) 
 & + (A(\bm u_{n+1}) (\varrho_{n+1} \bm u_{n+1}), \bm u_{n+1} ) \\
 & = (A(\bm u_{n+1}) (\varrho_{n+1} \bm u_{n+1}), \bm u_{n+1} )
 - \frac{1}{2} (A(\bm u_{n+1}) \varrho_{n+1}), |\bm u_{n+1}|^2) = 0 .
\end{split} 
\]  
This makes possible to rewrite the inequality (\ref{27}) as
\begin{equation}\label{28}
 \frac{1}{2} \left ( \frac{\varrho_{n+1} |\bm u_{n+1}|^2 - \varrho_{n} |\bm u_{n}|^2  }{\tau } , 1 \right ) 
 - (p(\varrho_{n+1}), \div \bm u_{n+1} ) \leq  0 . 
\end{equation} 

To estimate the second term on the left--hand side of the inequality (\ref{28}),
we apply the discrete analogue of the renormalized equation of continuity (\ref{10}).
Multiply the continuity equation (\ref{22}) by ${\displaystyle \frac{d \varPi}{d \varrho}  (\varrho_{n+1})}$:
\begin{equation}\label{29}
 \frac{\varrho_{n+1} - \varrho_{n} }{\tau}  \frac{d \varPi}{d \varrho} (\varrho_{n+1}) 
 + A(\bm u_{n+1}) \varrho_{n+1}  \frac{d \varPi }{d \varrho} (\varrho_{n+1}) = 0 . 
\end{equation} 
The following equality takes place:
\[
 \varPi(\varrho_{n+1}) - \varPi(\varrho_{n}) = \frac{d \varPi}{d \varrho} (\varrho_{n+1}) (\varrho_{n+1}-\varrho_{n})
 - \frac{1}{2} \frac{d^2 \varPi}{d \varrho^2} (\widetilde{\varrho}^{n+1}) (\varrho_{n+1}) (\varrho_{n+1}-\varrho_{n})^2 ,
\] 
where
\[
 \widetilde{\varrho}^{n+1} \in [ \min(\varrho_{n}, \varrho_{n+1}), \ \max(\varrho_{n}, \varrho_{n+1}) ] .
\] 

\begin{as}\label{as-2}
Assume that
\begin{equation}\label{30}
 \frac{d^2 \varPi}{d \varrho^2} (\varrho) \geq 0 , 
\end{equation} 
\end{as}

Under these natural assumptions, we get
\[
 \frac{d \varPi}{d \varrho} (\varrho_{n+1}) (\varrho_{n+1}-\varrho_{n}) \geq  
 \varPi(\varrho_{n+1}) - \varPi(\varrho_{n}) .
\]
For the second term in (\ref{29}), taking into account (\ref{8}), we have
\[
\begin{split}
 A(\bm u_{n+1}) \varrho_{n+1}  \frac{d \varPi }{d \varrho} (\varrho_{n+1}) 
 & =  \varrho_{n+1}  \frac{d \varPi }{d \varrho} (\varrho_{n+1}) \div \bm u_{n+1} + \bm u_{n+1} \grad \varPi (\varrho ^{n+1}) \\
 & = \div(\varPi(\varrho_{n+1}) \bm u_{n+1}) + p(\varrho_{n+1}) \div \bm u_{n+1} .
\end{split} 
\]
In view of this, integration of equation (\ref{29}) results in
\begin{equation}\label{31}
 \left (\frac{\varPi(\varrho_{n+1}) - \varPi(\varrho_{n})}{\tau} , 1 \right ) + (p(\varrho_{n+1}), \div \bm u_{n+1}) \leq 0 .  
\end{equation}
Combining (\ref{28}) and (\ref{31}), we obtain the inequality
\begin{equation}\label{32}
 \left (\frac{1}{2} \varrho_{n+1} |\bm u_{n+1}|^2 + \varPi(\varrho_{n+1}) , 1 \right ) 
 \leq \left (\frac{1}{2} \varrho_{n} |\bm u_{n}|^2 + \varPi(\varrho_{n}) , 1 \right ) .
\end{equation} 

Comparing (\ref{32}) with (\ref{12}), we can conclude that at the discrete level, instead of fulfillment
of the conservation law for the total energy, we observe a decrease of the energy. It should be noted that
this property has been established under the additional assumption (\ref{30}).
The result of our consideration can be expressed in the following statement.

\begin{pr}\label{p-1}
The fully implicit scheme (\ref{22}), (\ref{23}) produces
the approximate solution of the problem (\ref{19})--(\ref{21}) that satisfies 
the mass conservation law in the form (\ ref {24}) and
the momentum conservation law (\ref{25}). Moreover, if the assumptions \ref{as-1} and \ref{as-2} hold,
the estimate (\ref{32}) for the total mechanical energy is also fulfilled.
\end{pr}

\section{Decoupling schemes} 

A linearized scheme is used, where the solution at a new time level is evaluated by advective transport
taken from the previous time level.
Using such a linearization, an iterative process is constructed for the numerical implementation
of the fully implicit scheme. The approximate solution at the new time level is determined by sequential solving,
first, the linear problem of advection for the density and, secondly, the linear problem for the velocity.

\subsection{Linearized scheme }

We focus on the use of such time--stepping techniques that demonstrate the following properties:
\begin{itemize}
 \item the transition to a new time level is implemented by solving linear problems;
 \item splitting with respect to physical processes is employed, namely, the problems for
 the density and velocity are solved separately (with individual problems for the velocity components).
\end{itemize}
An example of the simplest decoupling scheme for the Euler equations system (\ref{19}), (\ref{20}) 
is the linearized scheme, where the advective transport involves the velocity from the previous time level.

Instead of (\ref{22}), (\ref{23}), we employ the scheme
\begin{equation}\label{33}
 \frac{\varrho_{n+1} - \varrho_{n} }{\tau} + A(\bm u_{n}) \varrho_{n+1}  = 0,
\end{equation} 
\begin{equation}\label{34}
 \frac{\varrho_{n+1} \bm u_{n+1} - \varrho_{n} \bm u_{n}  }{\tau } + 
 A(\bm u_{n}) (\varrho_{n+1} \bm u_{n+1}) + \grad p(\varrho_{n+1}) = 0, 
 \quad n = 0, 1, ..., N-1 .
\end{equation}
First, from the linear transport equation (\ref{33}), we evaluate the density at the new time level.
Next, from the linear decoupled system (\ref{34}) for the velocity components, we calculate
the velocity $\bm u_{n+1}$.

\begin{rem}
The system of equations (\ref{34}) with a given density is, in the general case, coupled
for individual cartesian velocity components. In the case, where all parts of the boundary 
of the computational domain $\partial \Omega$ are parallel to the axes of the cartesian coordinate system, 
the system of equations (\ref{34}) is decoupled and so, we can evaluate independently the individual components 
of the velocity.
\end{rem}

For the linearized scheme, the discrete analogs of the mass conservation law (see (\ref{24})) 
and the momentum conservation law (see (\ref{25})) hold.

\subsection{Iterative decoupling scheme} 

On the basis of the linearized scheme (\ref{33}), (\ref{34}),
it is possible to construct an iterative algorithm for the numerical implementation 
of the fully implicit scheme (\ref{22}), (\ref{23}).
The approximate solution for $\varrho_{n+1}$, $\bm u_{n+1}$ at the $k$--th iteration is denoted by
$\varrho_{n+1}^{k}$, $\bm u_{n+1}^{k}$, with the initial approximation from the previous time level:
\begin{equation}\label{35}
 \varrho_{n+1}^0 = \varrho_{n},
 \quad \bm u_{n+1}^0 = \bm u_{n} .
\end{equation}
Assume that the new approximation at the new time level is calculated, when the previous $K$ iterations have been done. 
Similarly to (\ref{33}), (\ref{34}), we use the system of equations:
\begin{equation}\label{36}
 \frac{\varrho_{n+1}^{k+1} - \varrho_{n} }{\tau} + A(\bm u_{n+1}^k) \varrho_{n+1}^{k+1}  = 0,
\end{equation} 
\begin{equation}\label{37}
 \frac{\varrho_{n+1}^{k+1} \bm u_{n+1}^{k+1} - \varrho_{n} \bm u_{n}  }{\tau } + 
 A(\bm u_{n+1}^{k}) (\varrho_{n+1}^{k+1} \bm u_{n+1}^{k+1}) + \grad p(\varrho_{n+1}^{k+1}) = 0, 
 \quad k = 0, 1, ..., K-1 ,
\end{equation} 
where
\begin{equation}\label{38}
 \varrho_{n+1} = \varrho_{n+1}^K,
 \quad \bm u_{n+1} = \bm u_{n+1}^K ,
 \quad n = 0, 1, ..., N-1 .
\end{equation}
Thus, at each iteration, we firstly solve the linear problem for the density and only then calculate the linear
problem for the velocity.

\section{Numerical results} 

The possibilities of the fully implicit scheme and decoupling schemes are illustrated by numerical results 
for a model two--dimensional problem with a density perturbation of an initially
resting fluid.

\subsection{Test problem} 

Here, we present the results of numerical solving a model problem obtained using
different time--stepping techniques.
The problem (\ref{1})--(\ref{4}) is considered in the square
\[
 \Omega = \{ \bm x \ | \ \bm x = (x_1, x_2), \quad - 5 <  x_1 < 5,
 \quad  \ - 5 <  x_2 < 5\} . 
\]
Assume that the dependence of the density on the pressure has the form ((\ref{9}) with $a = 1$, $\gamma = 1.4$.
We simulate the motion of the initially resting fluid ($\bm u^0(\bm x) = 0$ in (\ref{4}))
with the initial density (see (\ref{4})) specified in the form
\[
 \varrho^0(\bm x) = 1 + \alpha \exp(- \beta |\bm x |^2),
\] 
where $\alpha = 2$ and $\beta = 20$. 

\subsection{Fully implicit scheme} 

The problem is solved using the standard uniform triangulation $\Omega$ 
on  $M$ segments in each direction. The piecewise--linear finite elements $P_1$ are employed for discretization in space. 
To implement the fully implicit scheme (\ref{22}), (\ref{23}) 
for the nonlinear discrete problem at the new time level, the Newton method
with a direct solver is applied for the corresponding system of linear algebraic equations.
 
\begin{figure}
\begin{minipage}{0.43\linewidth}
\centering
\includegraphics[width=\linewidth]{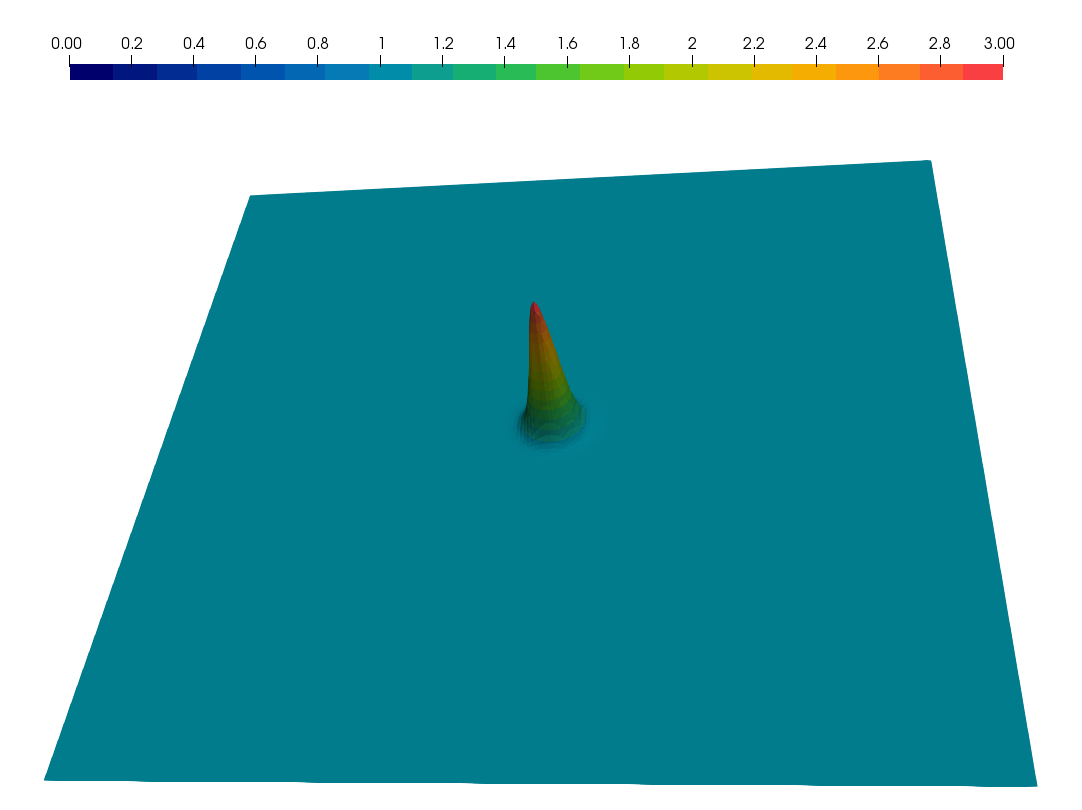}\\
$t=0$ \\
\includegraphics[width=\linewidth]{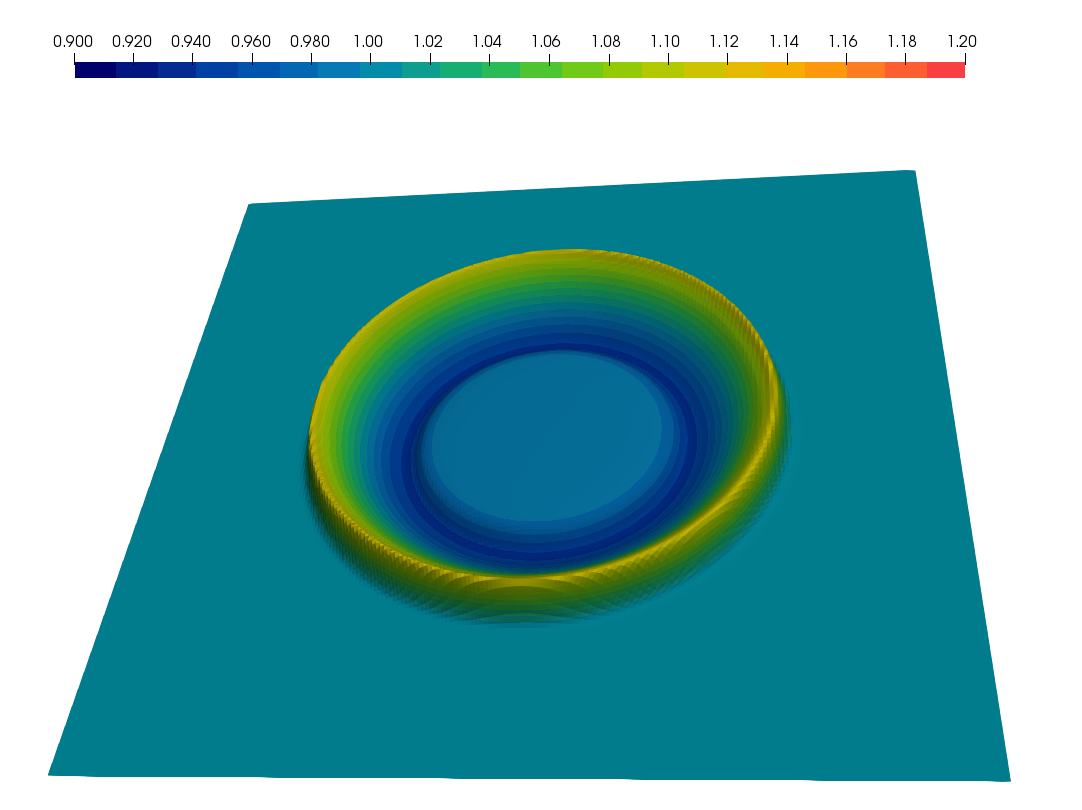}\\
$t=2$ \\
\includegraphics[width=\linewidth]{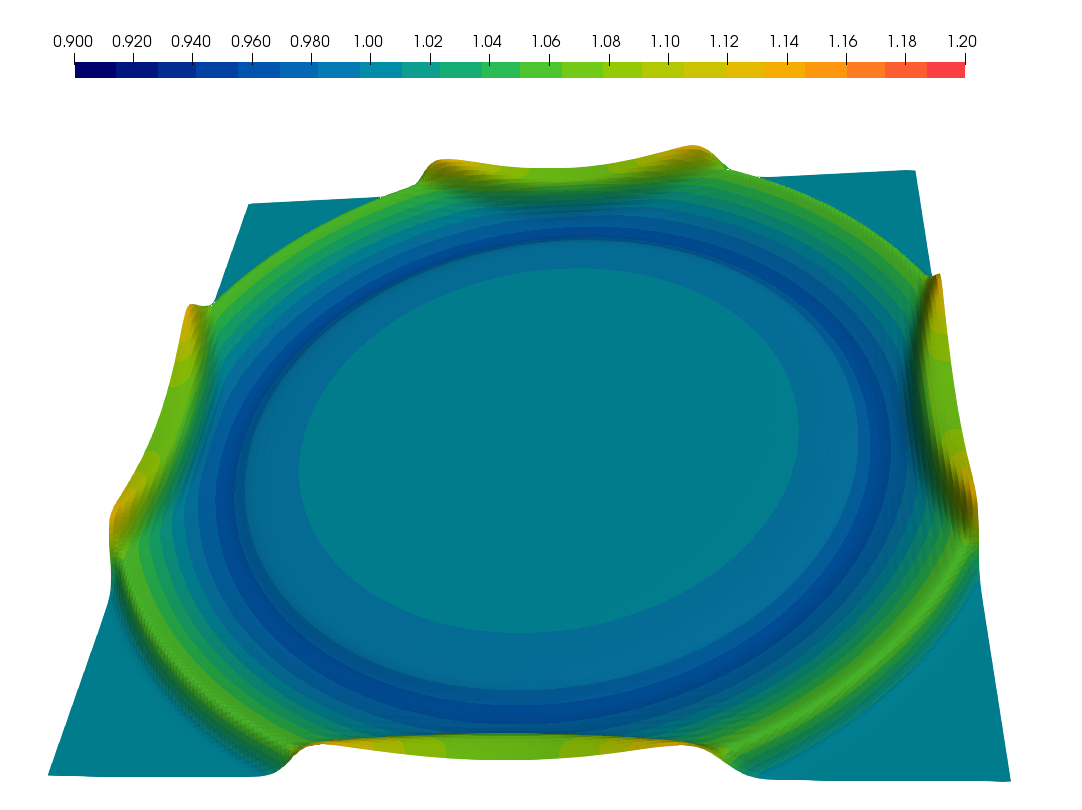}\\
$t=4$ 
\end{minipage}
\begin{minipage}{0.43\linewidth}
\centering
\includegraphics[width=\linewidth]{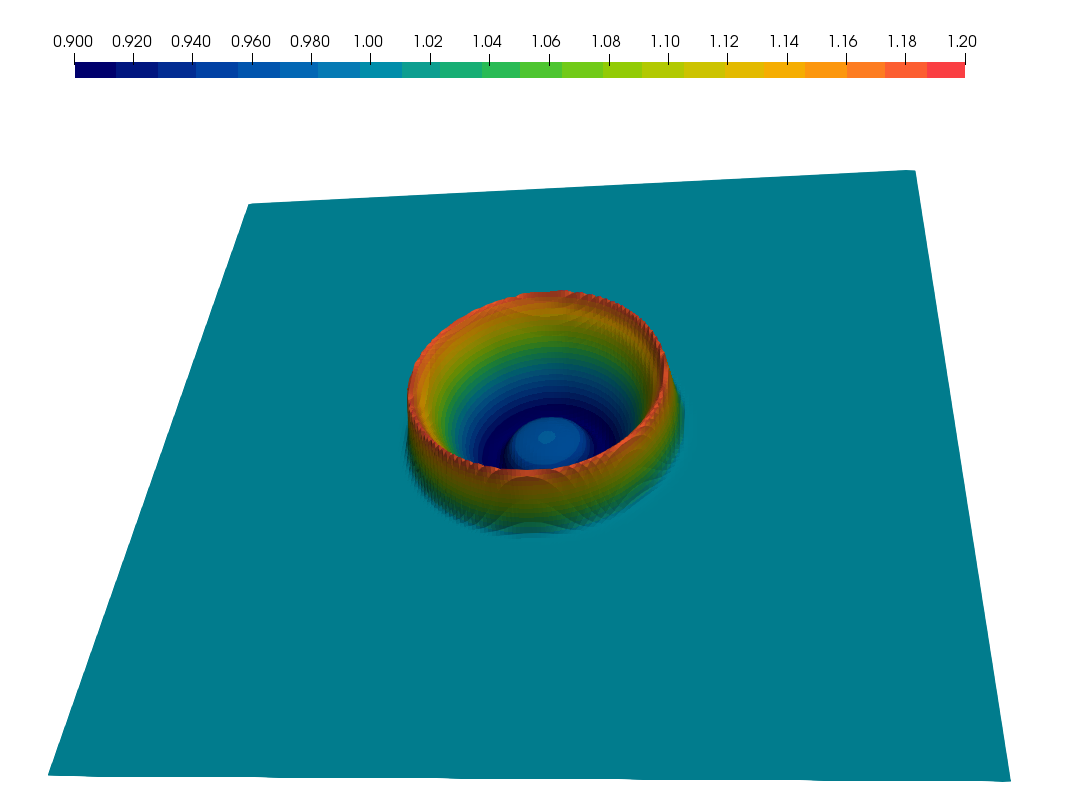}\\
$t=1$ \\
\includegraphics[width=\linewidth]{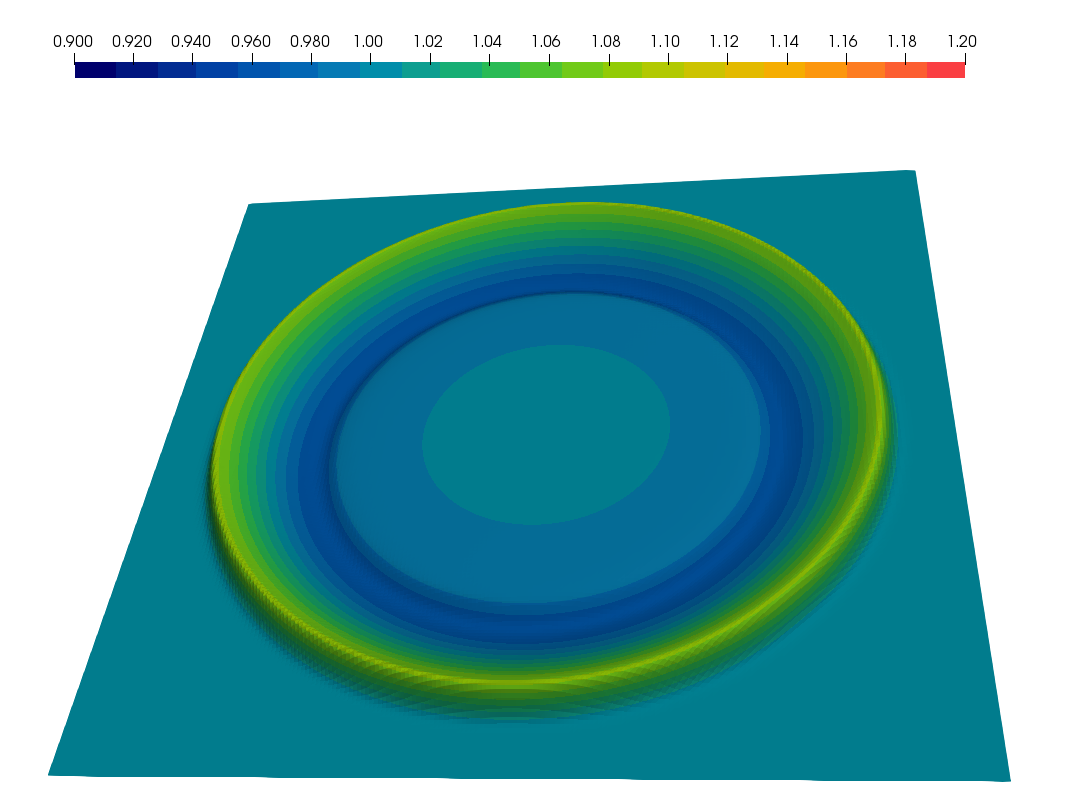}\\
$t=3$ \\
\includegraphics[width=\linewidth]{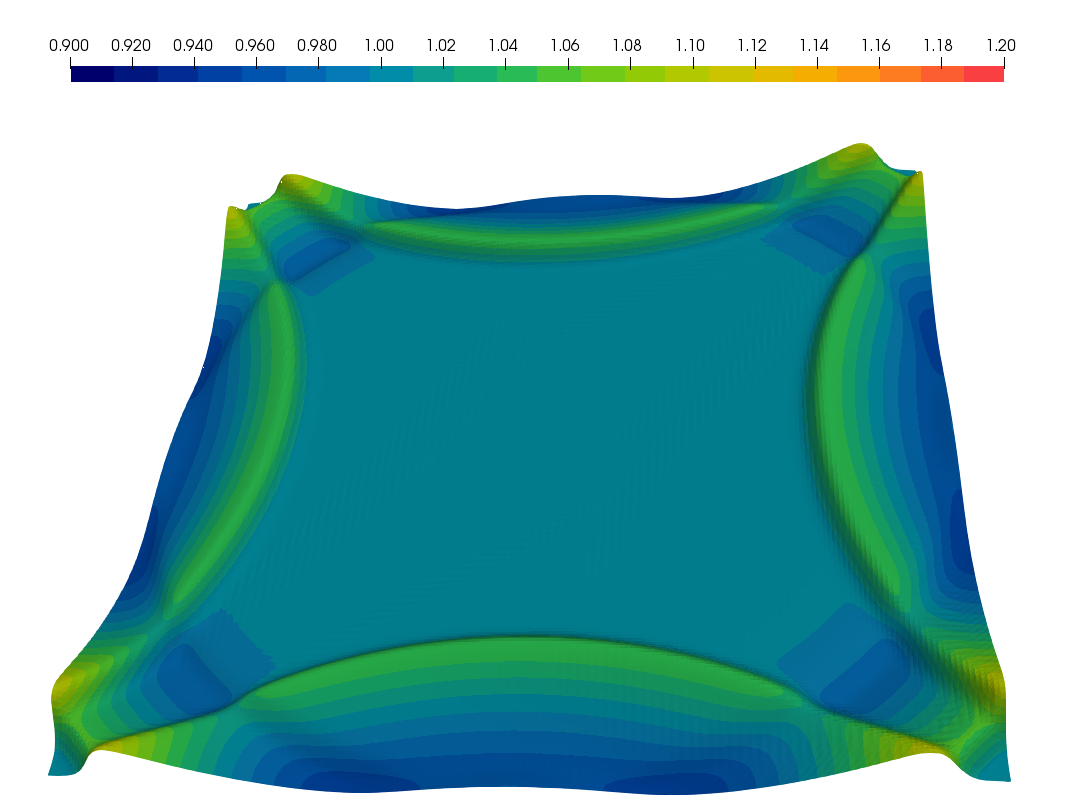}\\
$t=5$ 
\end{minipage}
\caption{The density at various time moments.}
\label{f-1}
\end{figure}

\begin{figure}
\centering
\includegraphics[width=\linewidth]{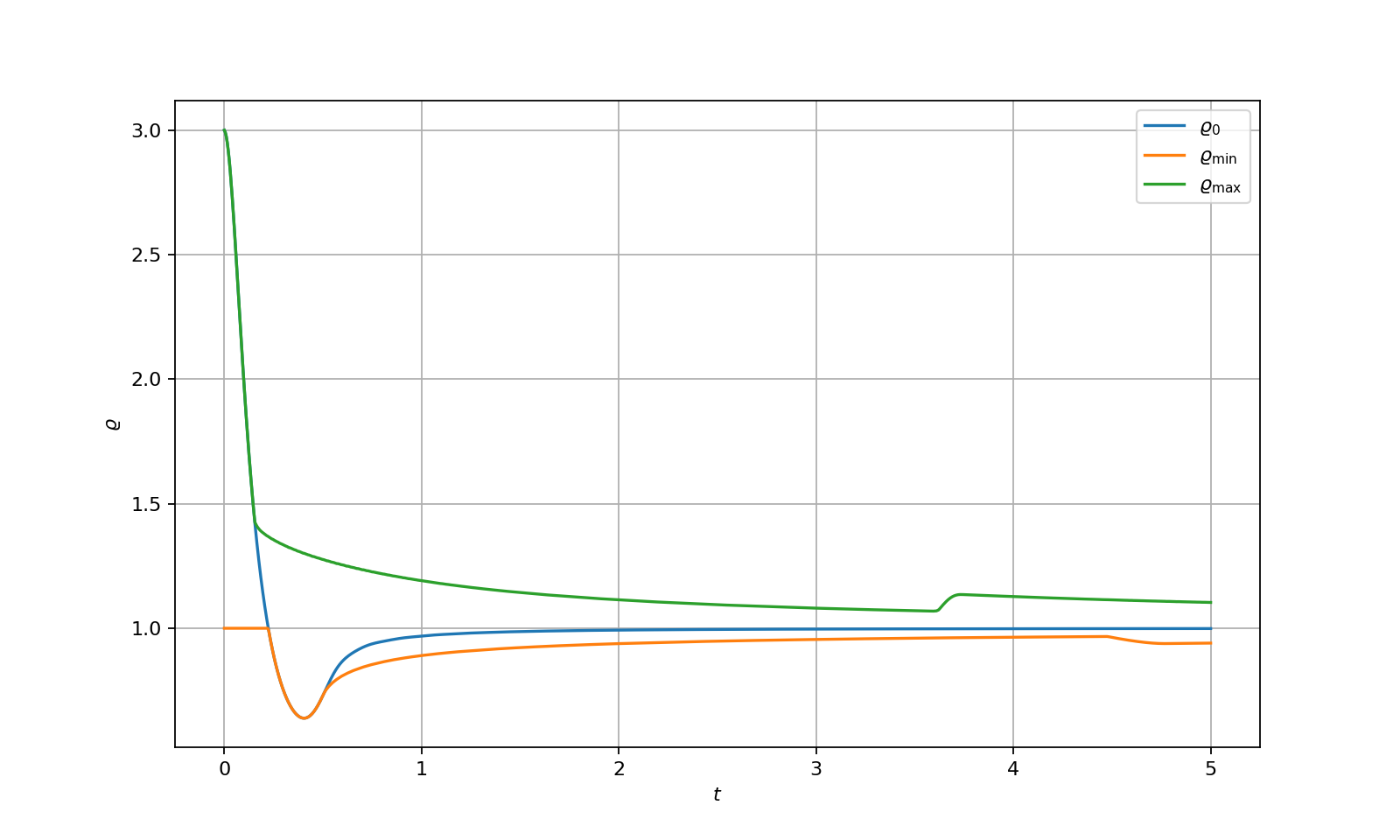}
\caption{Time--histories of the density (central, maximal and minimal values).}
\label{f-2}
\end{figure}

Time--evolution of the compressible fluid flow is shown in Fig.~\ref{f-1}, which presents the density 
at various time moments. In this calculation, we used the spatial grid with $M = 200$, the time step was $\tau = 0.005$. 
Time--histories of the density in the center of the computational domain ($\varrho_0$), maximum ($\varrho_{\max}$) 
and minimal ($\varrho_{\min}$) values of the density over the entire domain $\Omega$ are given in Fig.~\ref{f-2}.

Newton's iterative method for solving the discrete problem at each new time moment
converges very quickly (two or three iterations are enough). 
Table~\ref{t-1} demonstrates convergence of the iterative process for the first step in time.
Here, we present the relative error for the first three iterations for the
model problem obtained with $M = 200$ and various time steps.

\begin{table}[htp]
\centering
\caption{Convergence of Newton's method}
\label{t-1}
  \begin{tabular}{c|c|c|c}
  iteration & $\tau = 0.01$  & $\tau = 0.005$  & $\tau = 0.0025$ \\
  \hline
  1	&   2.219e-03  &  5.602e-04  &  1.404e-04  \\
  2	&   5.104e-08  &  9.100e-10  &  1.458e-11  \\
  3	&   5.566e-15  &  9.994e-15  &  1.788e-14  \\
  \end{tabular}
\end{table} 

\begin{figure}
\centering
\includegraphics[width=\linewidth]{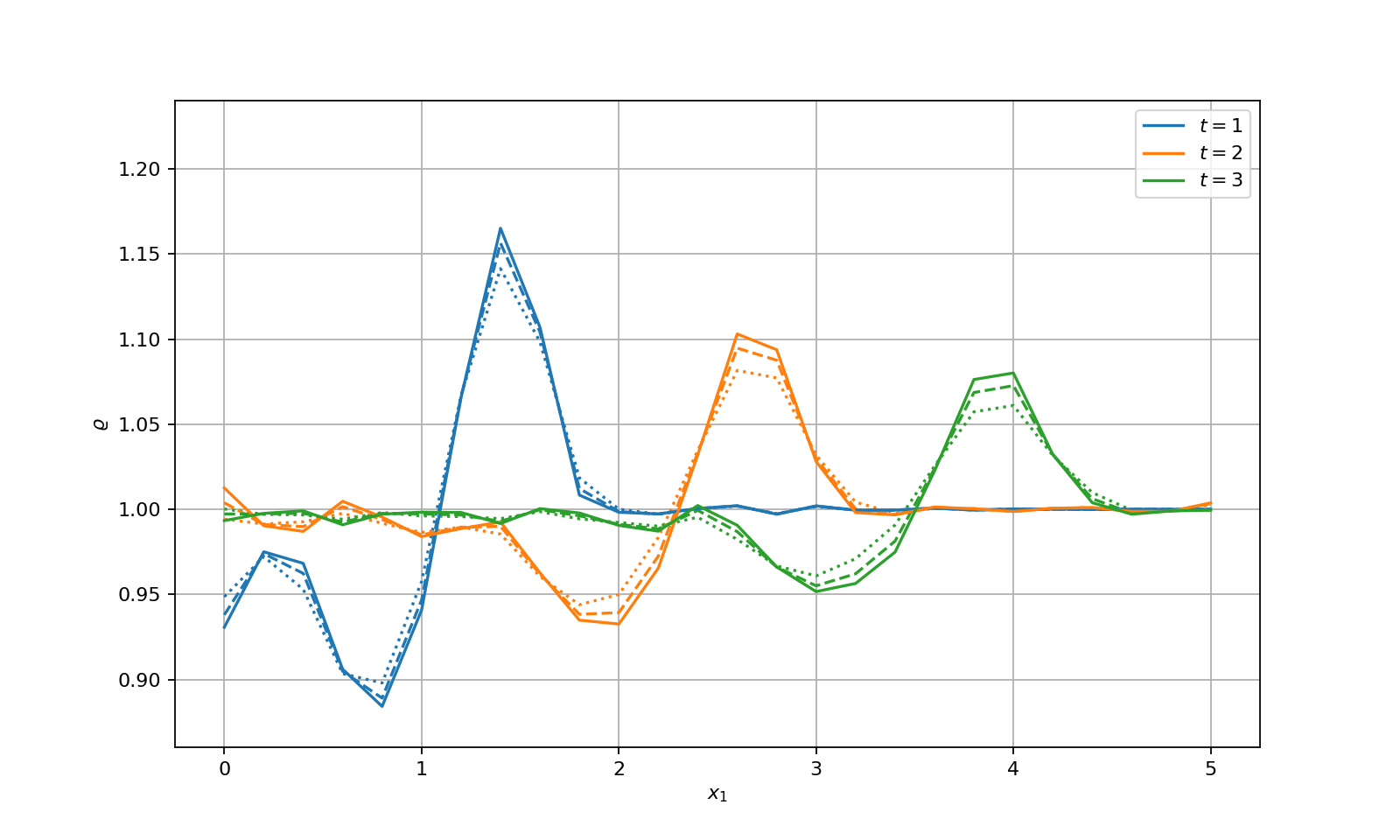}
\caption{The solution of the problem at various time moments calculated on the grid $M = 50$:
dotted line --- $\tau = 0.01$, dashed --- $\tau = 0.005$, solid --- $\tau = 0.0025$.}
\label{f-3}
\end{figure}

\begin{figure}
\centering
\includegraphics[width=\linewidth]{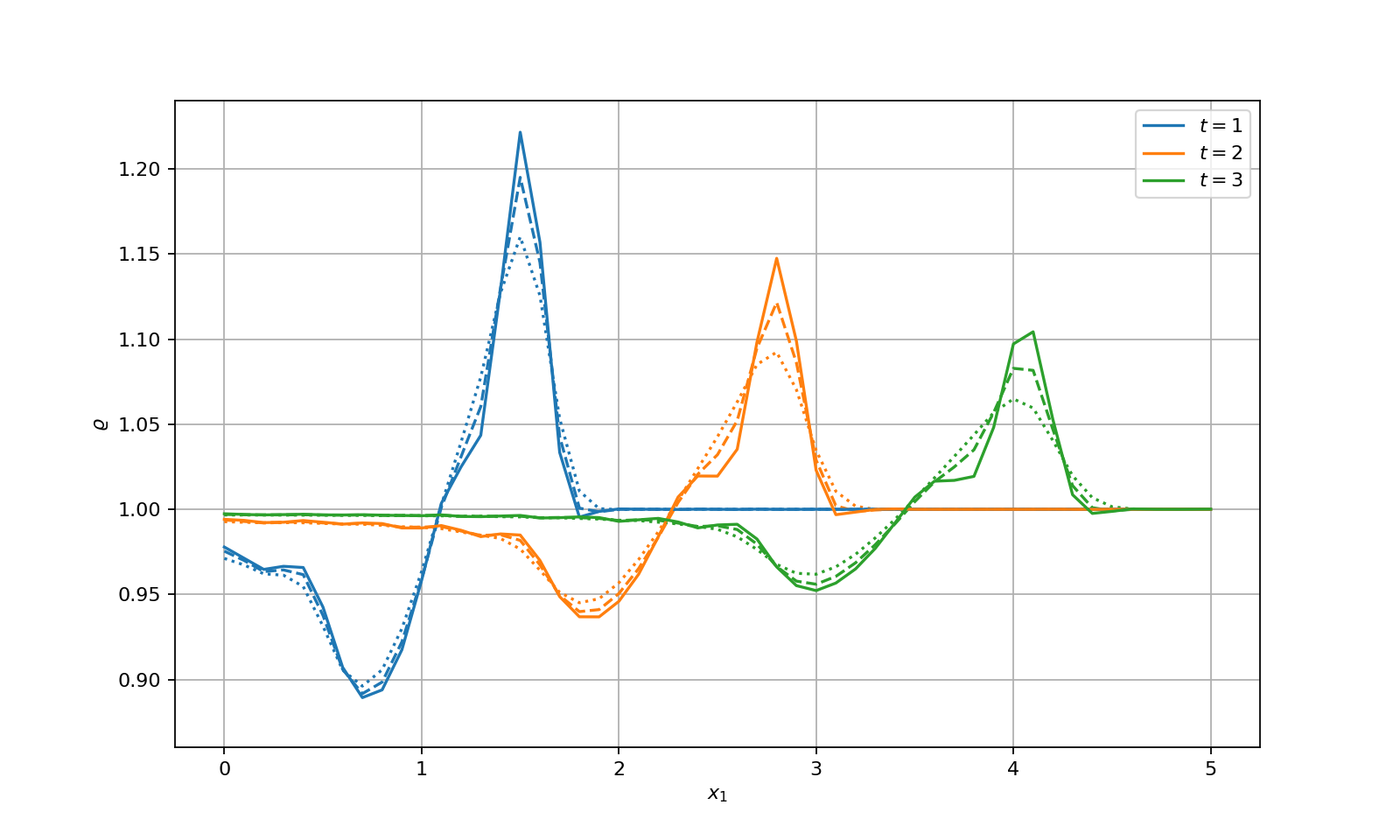}
\caption{The solution of the problem at various time moments calculated on the grid $M = 100$:
dotted line --- $\tau = 0.01$, dashed --- $\tau = 0.005$, solid --- $\tau = 0.0025$.}
\label{f-4}
\end{figure}

The accuracy of the approximate solution of the test problem will be illustrated by the data on the density 
in the section $x_2 = 0$. The solution calculated on the grid with $M = 50$ using various grids 
in time is shown in Fig.~\ref{f-3}.
Similar data for $M = 100$  and $M = 200$ are shown in Fig.~\ref{f-4},~\ref{5}, respectively.
It is easy to see a good accuracy in the reconstruction of the leading edge of the wave,
when the spatial grid is refined. Also, we observe the effect of smoothing, namely,
elimination of non--monotonicity with increasing of time step.

\begin{figure}
\centering
\includegraphics[width=\linewidth]{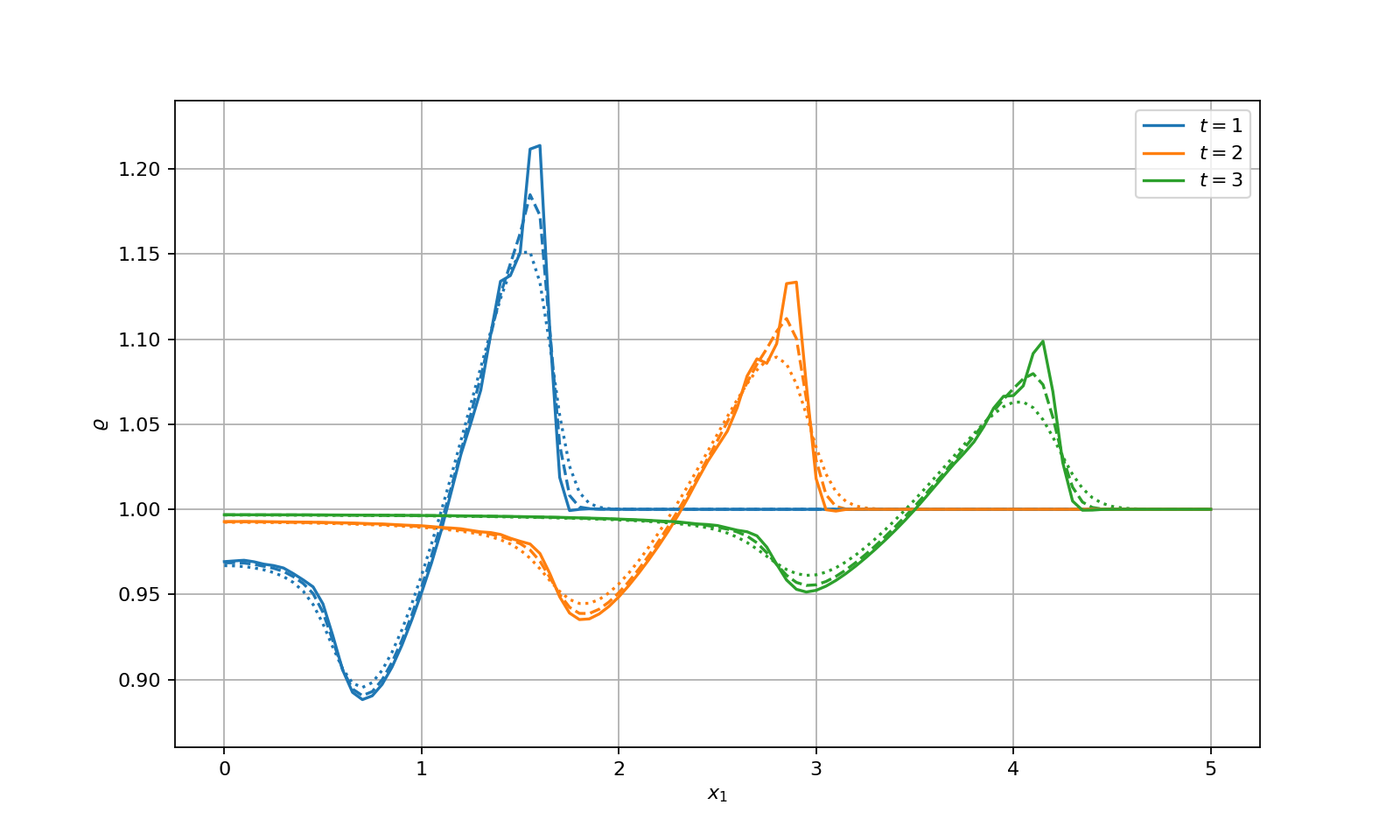}
\caption{The solution of the problem at various time moments calculated on the grid $M = 200$:
dotted line --- $\tau = 0.01$, dashed --- $\tau = 0.005$, solid --- $\tau = 0.0025$.}
\label{f-5}
\end{figure}

\subsection{Decoupling scheme} 

\begin{figure}
\centering
\includegraphics[width=\linewidth]{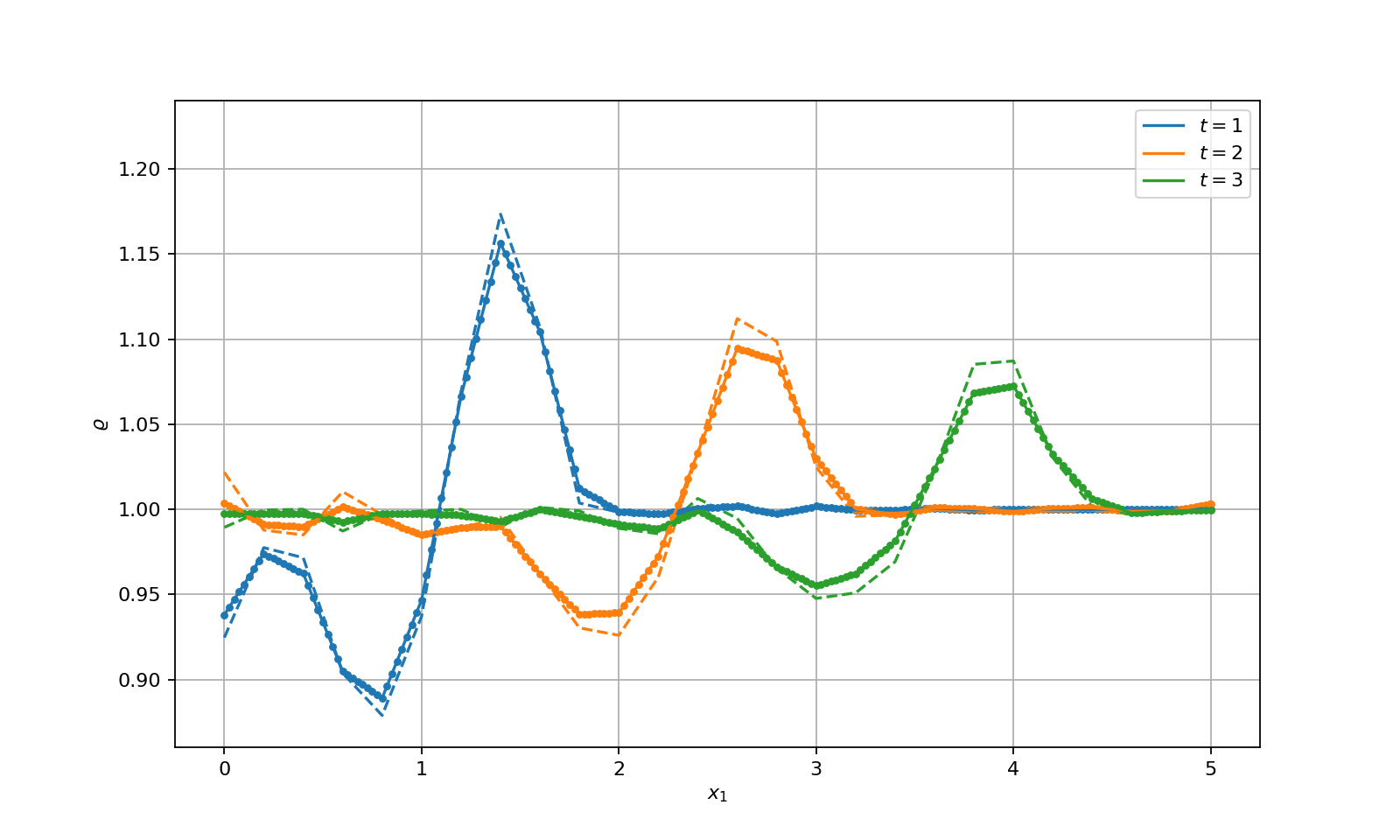}
\caption{The solution of the problem at various time moments calculated on the grid $M = 50$:
dashed line --- $K = 1$, dotted --- $K = 2$, solid --- $K = 5$.}
\label{f-6}
\end{figure}

\begin{figure}
\centering
\includegraphics[width=\linewidth]{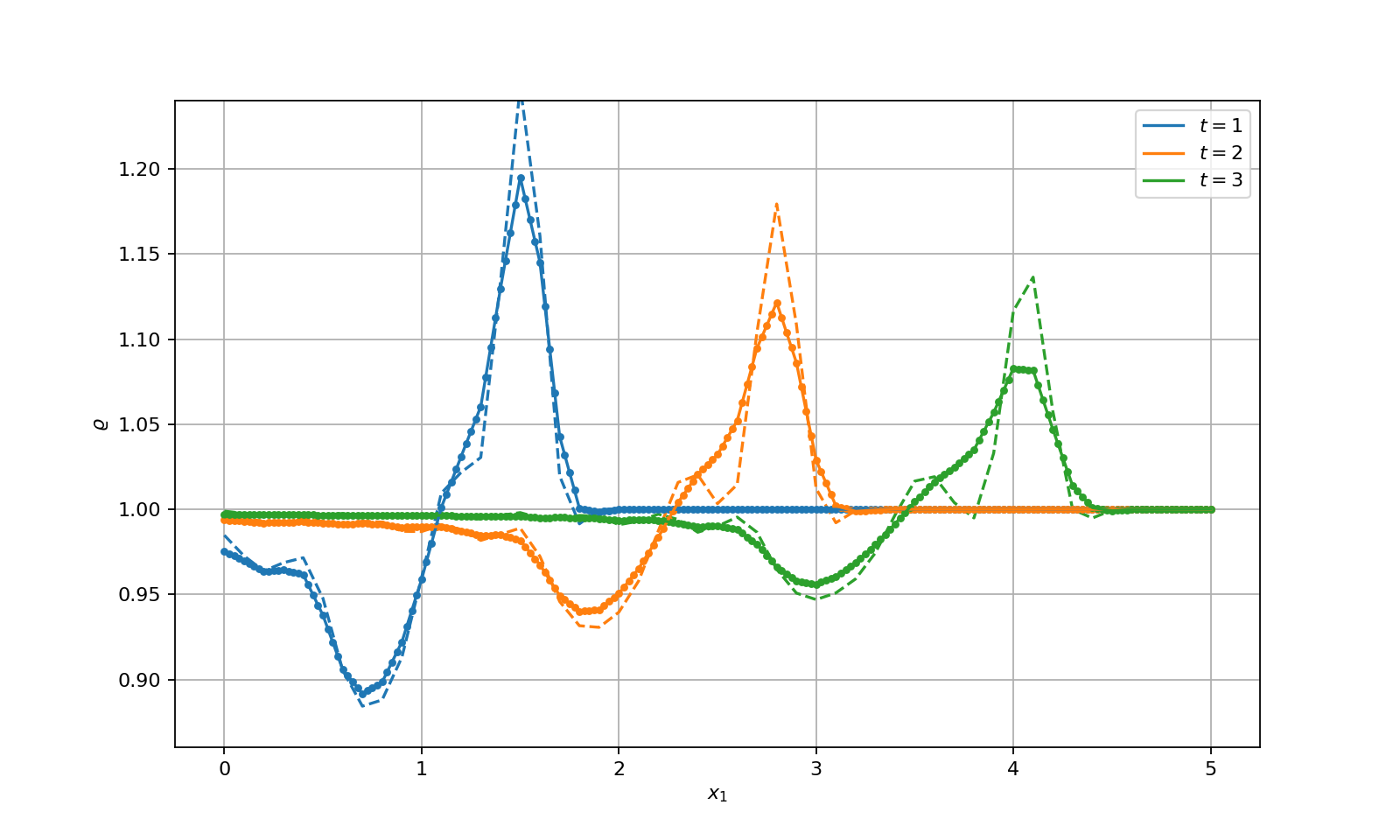}
\caption{The solution of the problem at various time moments calculated on the grid $M = 100$:
dashed line --- $K = 1$, dotted --- $K = 2$, solid --- $K = 5$.}
\label{f-7}
\end{figure}

\begin{figure}
\centering
\includegraphics[width=\linewidth]{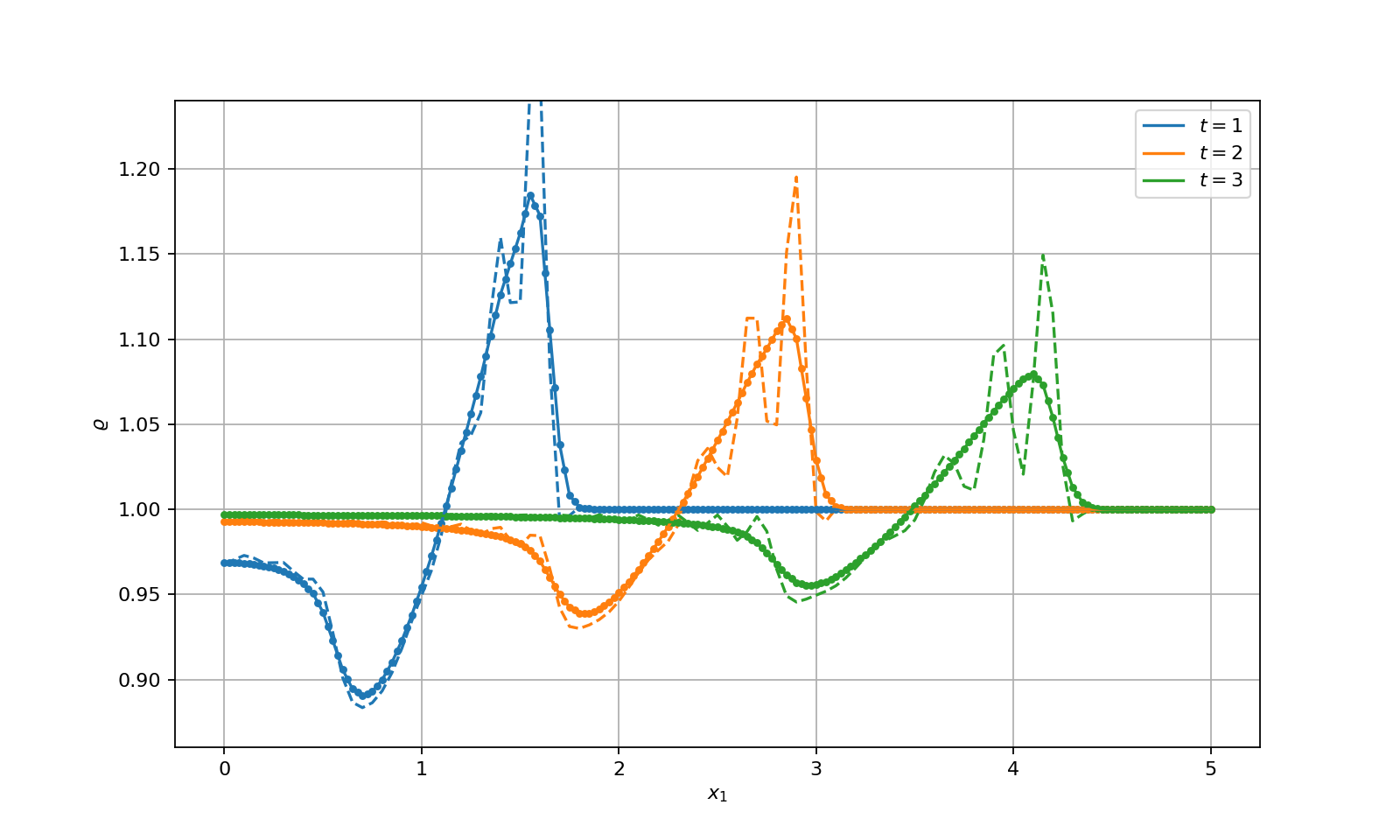}
\caption{The solution of the problem at various time moments calculated on the grid$M = 200$:
dashed line --- $K = 1$, dotted --- $K = 2$, solid --- $K = 5$.}
\label{f-8}
\end{figure}

In using the decoupling scheme (\ref{35})--(\ref{38}),
the greatest interest is related to the convergence rate of the iterative process.
The time step in the calculations was equal to $\tau = 0.005$.
We present numerical results for the model problem under consideration obtained on different grids in space. 
The dependence of the solution on the number of iterations for the grid with $M = 50$ is shown in Fig.~\ref{f-6}.
Figure~\ref{f-7} and \ref{f-8} presents similar data for grids with
$M = 100$ and $M = 200$, respectively.
It is easy to see see that on the finest grid (see Fig.~\ref{f-8}) the linearized scheme (\ref{34}), (\ref{34})) 
($K = 1$ in (\ref{35})--(\ref{38})) yields a substantially non--monotonic solution, which is monotonized 
on subsequent iterations.

The main conclusion of our study is the demonstration of high computational efficiency
of the iterative decoupling scheme (\ref{35})--(\ref{38}). Namely,
for the problems under consideration it is sufficient to do only
two iterations using (\ref{35})--(\ref{38}).

For the above time--stepping methods, the key point
is a violation of the conservation law for the total energy.
For the fully implicit scheme (\ref{22}), (\ref{23}), instead of conservation of the energy
(see the estimate (\ref{32})), decreasing of the total energy is observed.

\begin{figure}
\centering
\includegraphics[width=\linewidth]{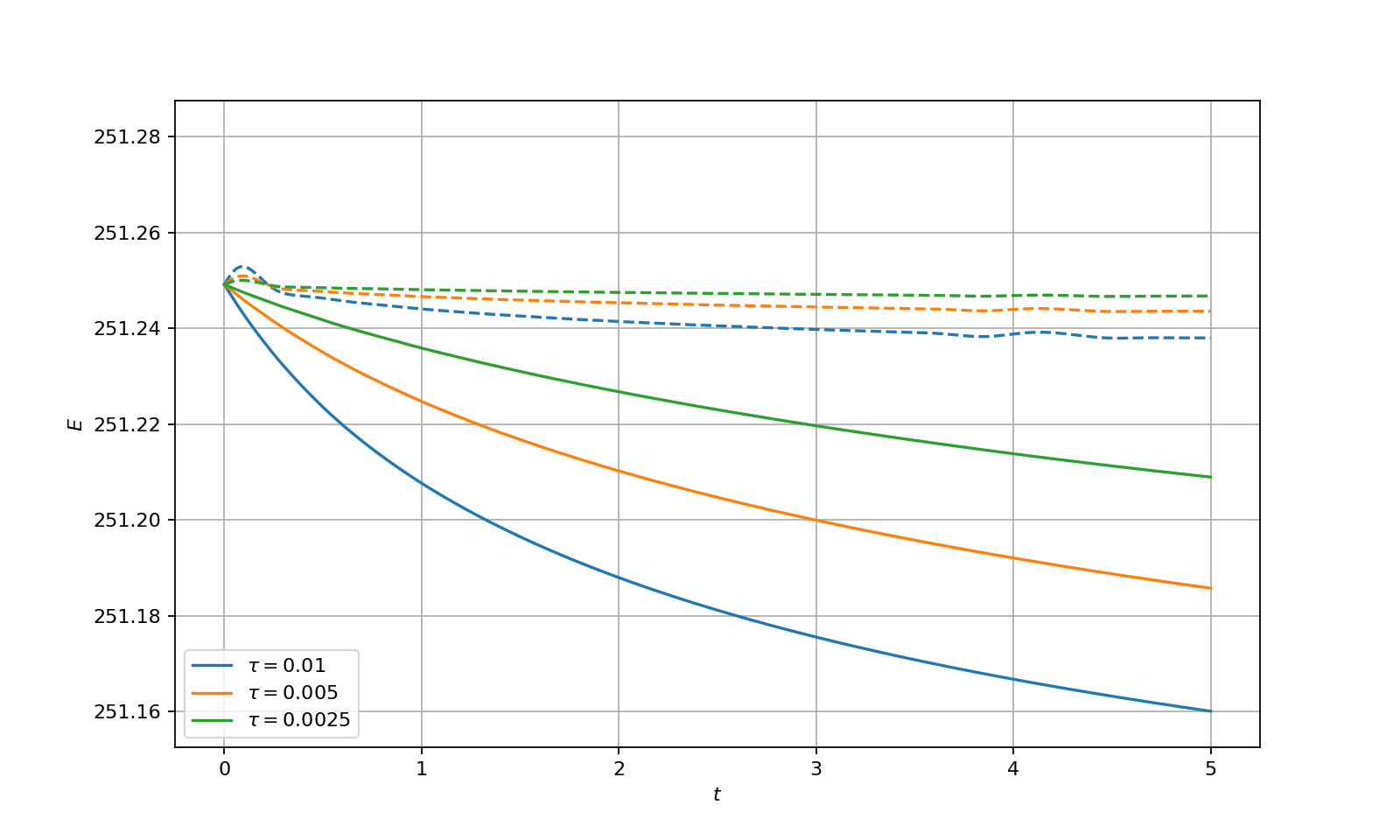}
\caption{Time--history of the total mechanical energy for various time steps obtained on the grid with $M = 50$:
dashed line --- $K = 1$, solid --- $K = 5$.}
\label{f-9}
\end{figure}

\begin{figure}
\centering
\includegraphics[width=\linewidth]{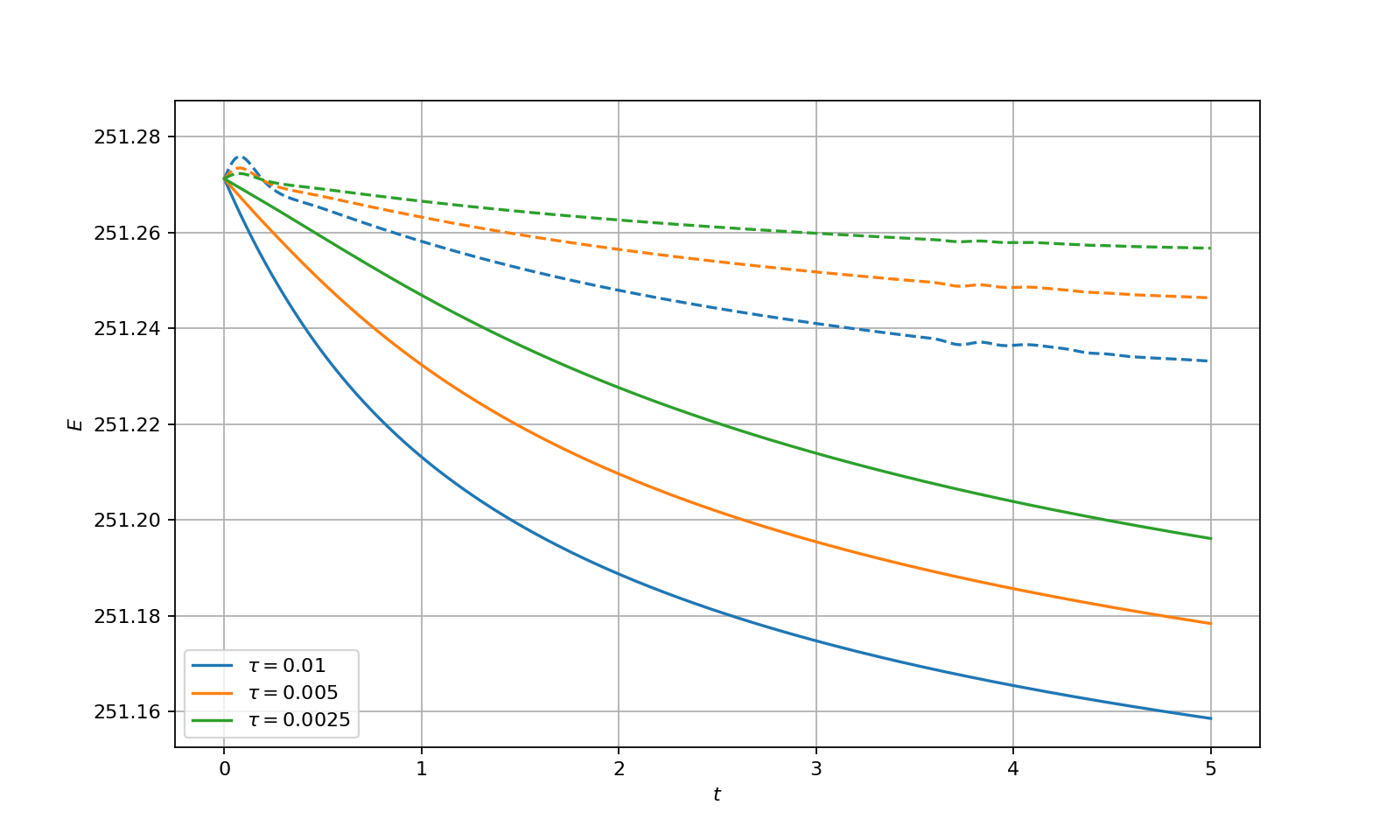}
\caption{Time--history of the total mechanical energy for various time steps obtained on the grid with $M = 100$:
dashed line --- $K = 1$, solid --- $K = 5$.}
\label{f-10}
\end{figure}

\begin{figure}
\centering
\includegraphics[width=\linewidth]{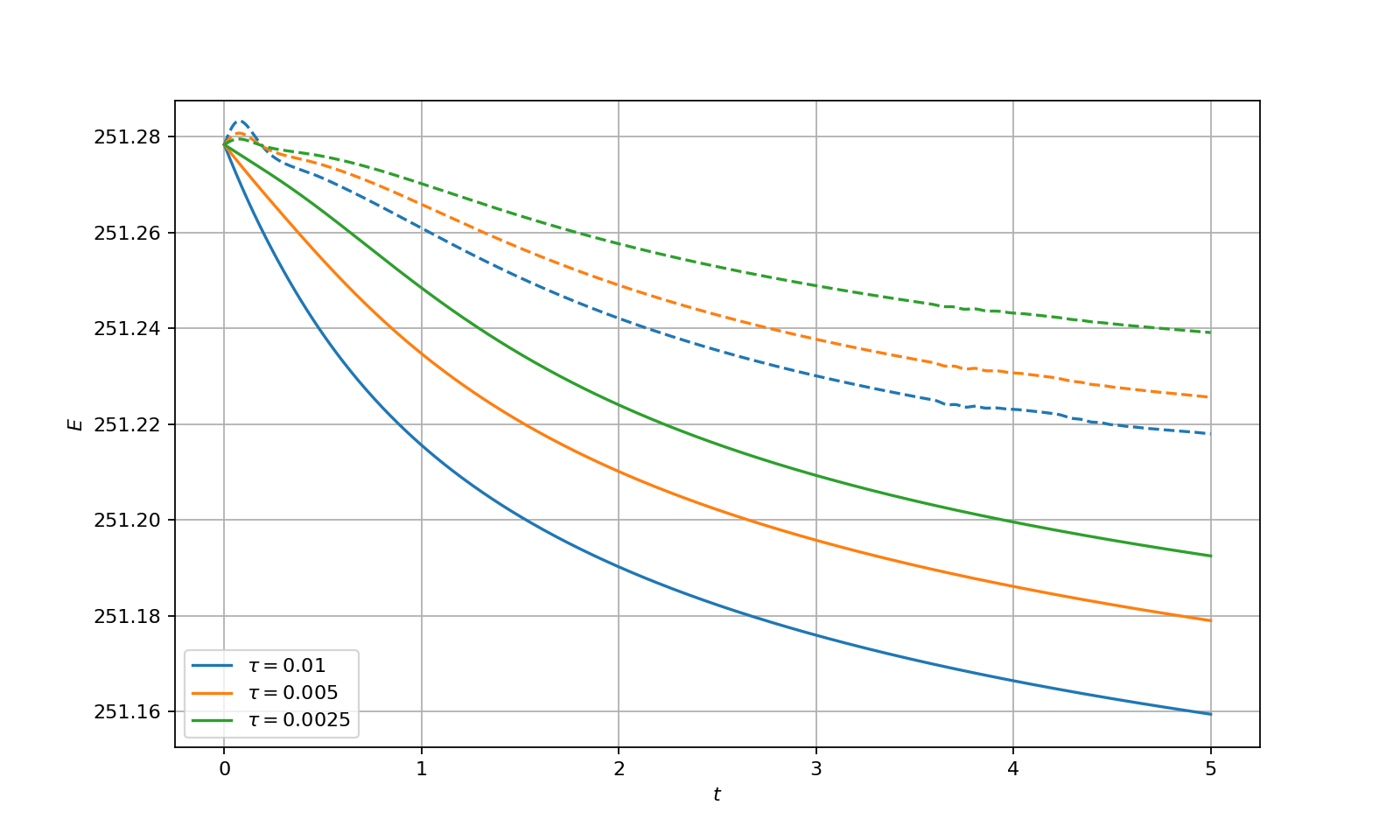}
\caption{Time--history of the total mechanical energy for various time steps obtained on the grid with $M = 200$:
dashed line --- $K = 1$, solid --- $K = 5$.}
\label{f-11}
\end{figure}

The dynamics of the total mechanical energy using a linearized scheme (\ref{33}), (\ref{34})
($K = 1$) and iterative decoupling schemes (\ref{35})--(\ref{38})) for $K = 5$
on various grids is shown in Fig.~\ref{f-9}--\ref{f-11}.
Here, according to (\ref{12}), $t_n$ is calculated at each time moment
\[
 E(t_n) =  \int_{\Omega}\left ( \frac{1}{2} \varrho_n |\bm u_n |^2 
 + \varPi(\varrho_n) \right ) d \bm x ,
 \quad n = 0, 1, ..., N . 
\]
For $K = 5$, the solution obtained using the decoupling scheme (\ref{35})--(\ref{38})
practically coincides with the solution derived from the fully implicit scheme (\ref{22}), (\ref{23}).
The above data indicate that the conservation law for the total energy
is satisfied with a good accuracy. Decreasing of the time step results in
increasing of the accuracy of the conservation law fulfillment.

\section*{Acknowledgements}

The publication was financially supported by the Ministry of Education and Science of 
the Russian Federation (the Agreement \# 02.a03.21.0008).

\clearpage


\end{document}